\documentclass[aps,prl,twocolumn,superscriptaddress]{revtex4}
\usepackage{makeidx}
\usepackage{graphicx}
\usepackage{amsmath}
\usepackage{amsfonts}
\usepackage{amssymb}
\usepackage{bm}
\usepackage{color}

\begin{document}

\title{Nernst-Ettingshausen effect in two-component electronic liquids}
\author{A.A.~Varlamov}
\affiliation{COHERENTIA-INFM, CNR, Viale del Politecnico 1, I-00133 Rome, Italy}
\author{A.V. Kavokin}
\affiliation{Dipartimento di Fisica, Universita' "Tor Vergata" via della Ricerca
Scientifica, 00133 Rome, Italy}
\affiliation{Physics and Astronomy School, University of Southhampton, Highfield,
Southampton, SO171BJ, UK}
\date{\today }

\begin{abstract}
A simple model describing the Nernst-Ettingshausen effect (NEE) in
two-component electronic liquids is formulated. The examples considered
include graphite, where the normal and Dirac fermions coexist,
superconductor in fluctuating regime, with coexisting Cooper pairs and
normal electrons, and the inter-stellar plasma of electrons and protons. We
give a general expression for the Nernst constant and show that the origin
of a giant NEE is in the strong dependence of the chemical potential on
temperature in all cases.
\end{abstract}

\maketitle

The Nernst-Ettingshausen effect (NEE) \cite{NE86} consists in induction of a
stationary electric field $E_{y}$ by the crossed magnetic field $H$ and
temperature gradient $\nabla T$ (see Fig.1) in the absence of electric
current. The Nernst constant $N=-\left( \frac{E_{y}}{H\nabla T}\right) $
varies drastically in different solid state systems, being extremely small ($%
N\sim $ $0.01-1\mu V\cdot K^{-1}T^{-1}$ ) in metals \cite{m1,m2}, but quite
large in semimetals and semiconductors ($N$ approaching $7000\mu V\cdot
K^{-1}T^{-1}$ in Bi \cite{BMK07}), very noticeable in the pseudogap state of
high temperature superconductors (HTSC) ($N\sim $ $1\mu V\cdot K^{-1}T^{-1}$%
) \cite{O03}, and, as it was recently discovered, it is even larger in
the fluctuation regime of a conventional superconductor $Nb_{0.15}Si_{0.85}$ ($%
N\sim 15\mu V\cdot K^{-1}T^{-1}$) \cite{P06}.

The existing theories of the Nernst effect are mainly based on the
semiclassical transport theory \cite{S48,Behnia2008} or Kubo formalism \cite%
{OU04}. We stress that the specific form of the transport equation differs
strongly from one system to another.  Also the form of the heat transfer
operator, fundamental for the Kubo approach,  strongly depends on the
character of interactions between the carriers. Moreover, in the existing
theories a single component electronic liquid has been addressed, which is
clearly not the case in semimetals \cite{BMK07}, superconductors above
critical temperature \cite{A07}\textbf{\ }and many other systems.

In this Letter, basing on the detailed current balance condition, we
propose a simple model where the Nernst coefficient is expressed in
terms of the chemical potential temperature derivative and
longitudinal conductivity. This approach uses the concept of
compensation of the drift current induced by crossed electric and
magnetic fields $j_{d}^{\left( x\right) }=qncE_{y}/H$, by a thermal
current induced by the temperature gradient. We apply it for
description of two-component electronic systems including
fluctuating superconductors, graphene/graphite, bismuth systems, and
the interstellar electron-proton\ plasma. Besides giving the general
formula for the Nernst coefficient in two-component electronic
liquids we predict the appearance of the carriers concentration
gradient along the temperature gradient and the longitudinal Nernst
effect for graphene.

\begin{figure}[b]
\includegraphics[width=.5\columnwidth]{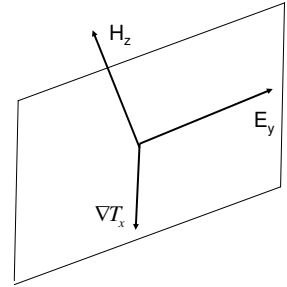}
\caption{Geometry of the Nernst-Ettingshausen effect. }
\end{figure}

Let us consider a conductor, placed in the magnetic field $\mathbf{H}$
oriented along z-axis, and\ subjected to the temperature gradient $%
\bigtriangledown T$ \ applied along x-axis (see Fig.1). There is no
electric currents flowing in the system, $j_{x}=j_{y}=0.$ Zero
current condition in the temperature gradient direction formally
yields:
\begin{equation}
\sigma _{xx}E_{x}+\sigma _{xy}E_{y}=0.  \label{current}
\end{equation}%
For an isotropic single component system with the carriers charge
$q$ and concentration $n$ the transversal (Hall) component of
conductivity is $\sigma _{xy}=-\sigma _{yx}=qnc/H$. Using the
constancy of the electrochemical potential $\mu \left( n,T\right)
-qE_{x}x$ \ along $x$-direction\ one can link the electric field
component$\ E_{x}$ to the temperature induced variation of the
chemichal potential: $qE_{x}=\nabla \mu \left( n,T\right) . $ The
electro-neutrality requirement in the case of one component system
provides the constancy of the charge concentration, so that
$qE_{x}=\left(
d\mu /dT\right) \nabla T.$ Substituting this relation to \ Eq. (\ref{current}%
) we obtain for the Nernst coefficient

\begin{equation}
N=\left( \frac{d\mu }{dT}\right) \frac{\sigma _{xx}}{q^{2}nc}.
\label{nernstdef}
\end{equation}

Eq. (\ref{nernstdef}) has been proposed for the first time in the
recent paper \cite{acceptedPRL}. Here we generalize it for the
two-component electronic systems and analyze the particular cases of
graphene/graphite, fluctuation superconductor and interstellar
plasma. Being extremely simple, Eq. (\ref{nernstdef}) works fairly
well. For instance, applying it to the degenerate $3D$ electron gas
in metals where $\sigma _{xx}=ne^{2}\tau /m$ \cite{Abrikosov}, while
$\tau $\ is the electronic mean free path time, $m$ is the electron
effective mass, and $\mu \left( T\right) =E_{F}-\pi
^{2}k_{B}^{2}T^{2}/\left( 12E_{F}\right) $ with $E_{F}$ being the
Fermi energy and $k_{B}$ being the Boltzmann constant, one can
obtain the familiar expression \cite{S48}:
\begin{equation*}
N_{\mathrm{e}}=-\frac{\pi ^{2}}{6}\left( \frac{k_{B}^{2}T}{E_{F}}\right)
\frac{\tau }{mc}.
\end{equation*}

As it was mentioned by Obraztsov \cite{O64} in some cases the Nernst
coefficient can be considerably renormalized by the effect of
magnetization currents developed in the sample when a magnetic field
is applied. For instance, this effect is negligible in the case of a
normal metal in non-quantizing fields ( by the parameter $\left(
a/\ell \right) ^{2},$ where $a$ is the lattice constant and $\ell $
is the electron mean free path) but affects significantly the value
of NEE coefficient in a fluctuating superconductor \cite{Ussishkin}.

Let us stress that the Eq. (\ref{nernstdef}) sheds light on the physical
origin of the strong variation of the Nernst constant in different solid
state systems. Roughly speaking, the NEE is strong in those systems where
the chemical potential strongly varies with temperature.

\bigskip

\textbf{\emph{Two component systems.}} In this case \ Eq.
(\ref{current}) needs to be applied for each type of carriers
separately. Indeed, in the stationary regime, there is no current in
$x$-direction for each particular type of the carriers because of
the absence of the circuit in the NEE geometry (see Figure 1). This
can be referred to as the detailed current balance condition. Having
in mind that the chemical potential for each type of carriers
depends on temperature and carrier concentration, so that

\begin{equation}
\nabla \mu _{1,2}\left( T,n\right) =\left( \frac{\partial \mu _{1,2}}{%
\partial T}\right) \nabla T+\left( \frac{\partial \mu _{1,2}}{\partial
n_{1,2}}\right) \nabla n_{1,2},  \label{mu12}
\end{equation}%
from the detailed current balance condition one readily obtains:
\begin{equation}
q_{1,2}^{2}n_{1,2}c\frac{E_{y}}{H}=\sigma _{xx}^{\left( 1,2\right) }\left[
\left( \frac{\partial \mu _{1,2}}{\partial T}\right) \nabla T+\left( \frac{%
\partial \mu _{1,2}}{\partial n_{1,2}}\right) \nabla n_{1,2}\right] .
\label{balance}
\end{equation}%
The total charge density remains constant what yields:
\begin{equation}
q_{1}\triangledown n_{1}+q_{2}\triangledown n_{2}=0  \label{concentr}
\end{equation}%
Finally, from Eqs. (\ref{balance})-(\ref{concentr}) one can express the
Nernst constant:

\begin{equation}
N\!=\!\frac{\left[ q_{1}\left( \frac{\partial \mu _{1}}{\partial T}\right)
\left( \frac{\partial \mu _{2}}{\partial n_{2}}\right) +q_{2}\left( \frac{%
\partial \mu _{2}}{\partial T}\right) \left( \frac{\partial \mu _{1}}{%
\partial n_{1}}\right) \right] }{c\left[ \left( \sigma _{xx}^{\left(
2\right) }\right) ^{-1}\left( \frac{\partial \mu _{1}}{\partial n_{1}}%
\right) \cdot q_{2}^{3}n_{2}+\left( \sigma _{xx}^{\left( 1\right) }\right)
^{-1}\left( \frac{\partial \mu _{2}}{\partial n_{2}}\right) q_{1}^{3}n_{1}%
\right] }.  \label{Nernst2compsigma}
\end{equation}

In the case of a single component electronic liquid Eq. (\ref%
{Nernst2compsigma}) naturally reduces to the Eq. (\ref{nernstdef}) obtained
above. Note also that the developed formalism applies both for bosonic and
fermionic charge carriers. In the rest of this Letter we apply the general
Eq. (\ref{Nernst2compsigma}) to the specific cases of semimetals, graphene,
fluctuation superconductor and interstellar plasma.

An interesting peculiarity of two-component electronic systems consists in
the appearance of the gradient of carrier concentration along the
temperature gradient. From Eq. (\ref{balance}) one readily obtains
\begin{equation*}
\nabla n_{1,2}=-\left( \frac{\partial \mu _{1,2}}{\partial n_{1,2}}\right)
^{-1}\nabla T\left[ \frac{q_{1,2}^{2}n_{1,2}c}{\sigma _{xx}^{\left(
1,2\right) }}N+\left( \frac{\partial \mu _{1,2}}{\partial T}\right) \right] .
\end{equation*}%
As it was mentioned above, in single-component systems the concentration
gradient is forbidden by the electric neutrality condition (\ref{concentr}) .

\bigskip

\textbf{\emph{Semimetals and graphene}}. In quasi-2D highly oriented
pyrolytic graphite, the 3D electrons having a parabolic spectrum coexist
with 2D holes having a linear (Dirac) energy spectrum \cite{kopelnatura,LK06}%
: $E\left( p\right) =\overline{c}p$ (here $p$ is the 2D quasi-momentum and $%
\overline{c}$ is the effective light speed.) The same situation is most
likely realized in Bismuth which exhibits quite similar to graphite
electronic characteristics \cite{KopPRB06}. From Eq. (\ref{Nernst2compsigma}%
) we readily obtain the expression for the Nernst constant as
\begin{equation}
N_{\left( \mathrm{sm}\right) }=\frac{1}{ce^{2}}\frac{\left[ \left( \frac{%
\partial \mu _{h}}{\partial T}\right) \left( \frac{\partial \mu _{e}}{%
\partial n_{e}}\right) -\left( \frac{\partial \mu _{e}}{\partial T}\right)
\left( \frac{\partial \mu _{h}}{\partial n_{h}}\right) \right] }{\left[
\left( \sigma _{xx}^{\left( h\right) }\right) ^{-1}n_{h}\left( \frac{%
\partial \mu _{e}}{\partial n_{e}}\right) -\left( \sigma _{xx}^{\left(
e\right) }\right) ^{-1}n_{e}\left( \frac{\partial \mu _{h}}{\partial n_{h}}%
\right) \right] }.  \label{Nernstgraph}
\end{equation}%
The conductivity of holes in the Boltzmann limit is given by $\sigma
_{xx}^{\left( h\right) }=\sigma _{xx}^{\left( 2Dh\right) }/a$, where \cite%
{Nomura}%
\begin{equation}
\sigma _{xx}^{\left( 2Dh\right) }=\frac{e^{2}}{\pi \hbar }l\sqrt{2\pi
n_{h}^{\left( 2D\right) }},  \label{condh}
\end{equation}%
with $l$ being the mean free path length for Dirac fermions, $n_{h}^{\left(
2D\right) }=an_{h}$. In the case of graphite, $a=3.35\mathring{A}$ is the
distance between neighboring graphene planes. The derivatives of the
chemical potential for the Boltzmann holes read as%
\begin{equation*}
\partial \mu _{h}/\partial n_{h}=\hbar \overline{c}a\sqrt{\pi /\left(
2n_{h}^{\left( 2D\right) }\right) },\qquad \partial \mu _{h}/\partial
T\approx -2k_{B}.
\end{equation*}%
Considering the $3D$ electrons as a degenerate Fermi gas, we obtain:
\begin{equation}
\frac{\partial \mu _{e}}{\partial n_{e}}=\frac{\hbar ^{2}}{m_{e}}\frac{\pi ^{%
\frac{4}{3}}}{\left( 3n_{e}\right) ^{\frac{1}{3}}},\quad \frac{\partial \mu
_{e}}{\partial T}=-\frac{\pi ^{2}k_{B}^{2}T}{2E_{F}}.  \label{dermun}
\end{equation}%
Substituting the Eqs. (\ref{condh}),(\ref{dermun},) into Eq. (\ref%
{Nernstgraph}) and taking the values of the parameters $\overline{c}%
=10^{8}cm/s,$ $m_{e}$ =$10^{-3}m_{0}$, from \cite{KopPRB06} with $m_{0}$
being the free electron mass, $n_{e}$=$10^{18}cm^{-3}$, $n_{h}^{\left(
2D\right) }$=$10^{11}cm^{-2}$ , $l=10^{-4}cm,\tau =1ps,$ we obtain $%
N_{\left( \mathrm{sm}\right) }$=-7.2$mV\cdot K^{-1}T$, at the liquid Helium
temperature, which is orders of magnitude larger than the Nernst constant in
metals. This value is very close to the value of $N$ reported in \cite{BMK07}
for Bismuth.

In graphene there is only one type of charge carriers: electrons having a
Dirac spectrum \cite{Aleiner2008}. Eq. (\ref{Nernstgraph}) reduces in this
case to
\begin{equation}
N_{\left( \mathrm{gr}\right) }=-\frac{2k_{B}l}{\pi \hbar c}\sqrt{\frac{2\pi
}{n_{e}^{\left( 2D\right) }}}  \label{graphene}
\end{equation}%
\ We expect the Nernst constant in graphene be of the same order as in
graphite and much larger than in conventional metals. This is consistent
with the recently published theory of thermoelectric and thermomagnetic
effects in graphene in the framework of the relativistic hydrodynamics \cite%
{Muller2008}.

Very interestingly, the conventional transverse Nernst effect we described
so far in graphene may be accompanied by an unconventional longitudinal NEE,
if the temperature gradient exceeds the critical value $\triangledown
T_{c}=N_{\left( \mathrm{gr}\right) }^{-1}\overline{c}/c.$ This critical
gradient corresponds to the drift current $j_{d}^{\left( x\right) }=-en_{e}%
\overline{c}$, which is the strongest drift current allowed in the system of
Dirac fermions having a fixed group velocity $\overline{c}$. If $%
\triangledown T>\triangledown T_{c}$, the drift current of Dirac fermions in
x-direction cannot fully compensate the thermo-current induced by the
temperature gradient $j_{th}^{\left( x\right) }=e\sigma _{xx}\left( \frac{%
\partial \mu _{e}}{\partial T}\right) \triangledown T$. \ In this regime, an
additional electric field parallel to the temperature gradient appears $%
E_{x}=\frac{1}{e}\left( \frac{\partial \mu _{e}}{\partial T}\right) \left(
\triangledown T-\triangledown T_{c}\right) $ which leads to the appearance
of the longitudinal NEE.
\begin{figure}[b]
%\vspace{30mm}
\includegraphics[width=.8\columnwidth]{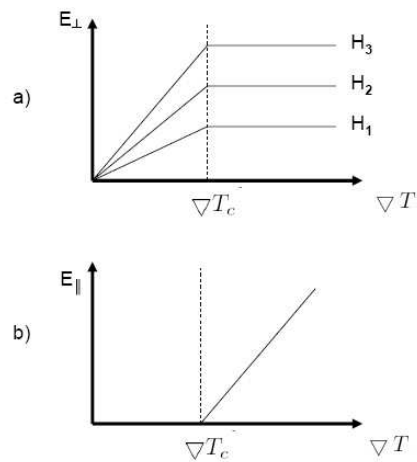}
\caption{Transverse (a) and longitudinal (b) electric fields as functions of
the temperature gradient and magnetic field ($H_{1}<H_{2}<H_{3}$) in
graphene. The vertical dashed (dotted) lines show the critical temperature
gradient in graphene.}
\end{figure}

Figure 2 shows schematically the transverse and longitudinal
electric field behavior as a function of the temperature gradient
for different values of magnetic field in graphene. One can see that
when the temperature gradient reaches its critical value, the
transverse NEE saturates, while the longitudinal NEE appears. This
threshold-like behavior of the Nernst fields is a signature of a
linear dispersion of the charge carriers.

\textbf{\emph{Fluctuating superconductor}}. As it is well known, the
NEE cannot be observed in type I superconductors due to the Meissner
effect, which does not allow magnetic field to penetrate in the
bulk. On the other hand, it is strong in type II superconductors in
the absence of pinning \cite{NII} due to the specific mechanism of
the entropy transport in process of vortex motion. Special attention
has been attracted recently by the giant NEE observed in the
pseudogap state of the underdoped phases of HTSC \cite{O03}, which
motivated speculations \cite{Anderson} about the possibility of
existence of some specific vortices and anti-vortices there or the
special role of the phase fluctuations \cite{Vis}. Finally, very
recently the giant NEE was found also in the wide range of
temperatures in a conventional disordered superconductor
$Nb_{x}Si_{1-x}$ \cite{P06} what has been successfully explained in
the frameworks of both phenomenological and microscopic \
fluctuation theories \cite{Ussishkin,acceptedPRL,Fink}. The
approach, based on Eq. (\ref{nernstdef}), allows not only to get in
a simple way the correct temperature dependence of the fluctuation
NEE coefficient but also to catch the reason of its giant magnitude.

In a superconductor being in the fluctuating regime Cooper pairs (cp)
coexist with normal electrons (e). Having in mind the double charge of the
Cooper pair, Eq. (\ref{Nernst2compsigma}) becomes

\begin{equation*}
N_{sc}=\frac{1}{ce^{2}}\frac{\left[ \left( \frac{\partial \mu _{e}}{\partial
T}\right) \left( \frac{\partial \mu _{cp}}{\partial n_{cp}}\right) +2\left(
\frac{\partial \mu _{cp}}{\partial T}\right) \left( \frac{\partial \mu _{e}}{%
\partial n_{e}}\right) \right] }{\left[ 8\left( \sigma _{xx}^{\left(
cp\right) }\right) ^{-1}\left( \frac{\partial \mu _{e}}{\partial n_{e}}%
\right) n_{cp}+\left( \sigma _{xx}^{\left( e\right) }\right) ^{-1}\left(
\frac{\partial \mu _{cp}}{\partial n_{cp}}\right) n_{e}\right] }.
\end{equation*}%
First of all let us mention that the value of chemical potential of
the gas of fluctuating Cooper pairs is defined by their `` binding
energy'' \cite{LV05} taken with the opposite sign: $\mu
_{cp}=-k_{B}\left( T-T_{c}\right) .$ Hence the terms containing
$\partial \mu _{cp}/\partial n_{cp}=0$ disappear.

In order to analyze both 2D and 3D cases simultaneously one can
evaluate the value of fluctuation contribution to the Nernst
coefficient on the example of a layered superconductor in the
vicinity of critical temperature and in the limit of weak magnetic
fields. The values of paraconductivity $\sigma _{xx}^{\left(
cp\right) }=e^{2}/\left( 16\hbar s\sqrt{\epsilon \left( \epsilon
+\epsilon _{\mathrm{cr}}\right) }\right) $ and fluctuation Cooper
pairs concentration ( up to the logarithmic accuracy) $n_{cp}\sim
k_{B}T_{c}/2\pi \mathcal{D}s\hbar $ in this case are available in
\cite{LV05}
(here $\epsilon $ is reduced temperature $(T-T_{c})/T_{c}$, $\epsilon _{%
\mathrm{cr}}$ is its crossover between 2D and 3D regimes value, $s$ is the
interlayer distance, and $\mathcal{D}$ is the in-plane diffusion
coefficient). In accordance with  \cite{UD92} we find
\begin{equation}
N_{\mathrm{fl}}=\left( \frac{\partial \mu _{cp}}{\partial T}\right) \frac{%
\sigma _{xx}^{\left( cp\right) }}{4n_{cp}ce^{2}}\sim -\frac{\pi \mathcal{D}}{%
32cT_{c}}\frac{1}{\sqrt{\epsilon \left( \epsilon +\epsilon _{\mathrm{cr}%
}\right) }}.  \label{zh}
\end{equation}%
The effect of fluctuation diamagnetic currents in this regime results in the
reduction of the Nernst constant by a factor of 3, according to  \cite%
{Ussishkin} , so that
\begin{equation*}
\widetilde{N}_{\mathrm{fl}}\sim -\frac{\pi \mathcal{D}}{96cT_{c}}\left\{
\begin{array}{cc}
\frac{T_{c}}{T_{c}-T}, & D=2 \\
\frac{1}{\epsilon _{\mathrm{cr}}}\sqrt{\frac{T_{c}}{T_{c}-T},} & D=3%
\end{array}%
\right. .
\end{equation*}

This formula in its $2D$ limit can be applied for description of the
experimental results obtained in thin films of the $Nb_{0.15}Si_{0.85}$ \cite%
{P06}. Substituting $T_{c}=0.38K$ and $\mathcal{D}=0.143cm^{2}/s$ we obtain $%
\widetilde{N}_{\mathrm{fl}}^{\left( 2\right) }\sim -T_{c}/\left(
T-T_{c}\right) \mu V\cdot T^{-1}K^{-1}$ , what corresponds well to
the experimental findings and is three orders of magnitude more than
the value of the Nernst coefficient in typical metals.

Hence one can conclude that the giant Nernst effect in fluctuating
superconductors comes from the strong dependence of the fluctuation Cooper
pairs chemical potential on temperature.

\bigskip

\textbf{\emph{Interstellar plasma}}. Consider the intergalactic medium which
is mostly ionized hydrogen, i.e. a plasma consisting of equal numbers of
electrons and protons, which exists at a density of 10 to 100 times the
average density of the Universe (10 to 100 hydrogen atoms per cubic meter)
\cite{plasma}. Both types of carriers can be described as classical gases
obeying the Boltzmann statistics. Since the plasma is electrically neutral,
one readily obtains from Eq. (\ref{Nernst2compsigma})%
\begin{equation*}
N_{\mathrm{isp}}=-3k_{B}\tau _{e}\tau _{p}\frac{m_{p}-m_{e}}{2c\left[
m_{p}^{2}\tau _{e}-m_{e}^{2}\tau _{p}\right] }\approx -\frac{3k_{B}\tau _{p}%
}{2cm_{p}},
\end{equation*}%
where $m_{e}\ $and $m_{p}$ are electron and proton masses, respectively. The
plasma temperature is thought to be quite high by terrestrial standards: it
heats up to $10^{5}K$ to $10^{7}K$, which is high enough for the bound
electrons to escape from the hydrogen nuclei upon collisions and corresponds
to speeds of the order of $10^{7}-10^{8}cm/s.$ Using the Rutherford formula
for the scattering cross-section with these data one can evaluate the
scattering time as $\tau _{p}=10^{2}-10^{3}s,$ which results in the huge NEE
coefficient of the order $N_{\mathrm{isp}}=10^{5}-10^{6}\mu V\cdot K^{-1}T$
! Observation of a giant NEE in the interstellar plasma could be an
important challenge for the experimental astrophysics.

In conclusion, we have formulated a unified approach to the
Nernst-Ettingshausen effect allowing for comparative analysis of the Nernst
constant in a wide range of systems. We expect the giant NEE to take place
in the charged systems where the chemical potential strongly depends on
temperature. We predict the huge values of the Nernst constant in graphene,
graphite and the electron-proton plasma and confirm the values of the Nernst
constant in metals, Bismuth and superconductors above the critical
temperature. The proposed approach could be the key for understanding of the
giant NEE in pseudogap phase of HTSC. We predict an existence of the
critical temperature gradient in graphene above which the unconventional
longitudinal NEE develops.

We are grateful to B.L.Altshuler, L. Falkovski, Y.Kopelevich and
I.Luk'yanchuk for valuable discussion.


\begin{references}
\bibitem{com} We stress, that the factor 3 of Nernst coefficient reduction valid only in the considered regime.
A detailed discussion on the magnetization current effects in
fluctuation superconductors can be found in Ref. 11.
\end{references}
\end{document}